# Social Connection Induces Cultural Contraction:
# Evidence from Hyperbolic Embeddings of Social and Semantic Networks

Lingfei Wu[a,b,c,d,e], Linzhuo Li[a,b,c] and James Evans[a,b,c,1]

[a] Department of Sociology, University of Chicago, 1126 E 59th St, Chicago, IL 60637, USA
[b] Knowledge Lab, University of Chicago, 5735 South Ellis Avenue, Chicago, IL 60637, USA
[c] Computation Institute, University of Chicago, 5735 South Ellis Avenue, Chicago, IL 60637, USA
[d] School of Journalism and Communication, Nanjing University, 22 Hankou Rd, Nanjing, China, 210008
[e] Tencent Research Institute, North 4th Ring West Road, Hai Dian District, Beijing, P.R. China 100080

Corresponding Author: Email: jevans@uchicago.edu

**Abstract**

Research has repeatedly demonstrated the influence of social connection and communication on convergence in cultural tastes, opinions and ideas. Here we review recent studies and consider the implications of social connection on cultural, epistemological and ideological contraction, then formalize these intuitions within the language of information theory. To systematically examine connectivity and cultural diversity, we introduce new methods of manifold learning to map both social networks and topic combinations into comparable, two-dimensional hyperbolic spaces or Poincaré disks, which represent both hierarchy and diversity within a system. On a Poincaré disk, radius from center traces the position of an actor in a social hierarchy or an idea in a cultural hierarchy. The angle of the disk required to inscribe connected actors or ideas captures their diversity. Using this method in the epistemic culture of 21$^{st}$ Century physics, we empirically demonstrate that denser university collaborations systematically contract the space of topics discussed and ideas investigated more than shared topics drive collaboration, despite the extreme commitments academic physicists make to research programs over the course of their careers. Dense connections unleash flows of communication that contract otherwise fragmented semantic spaces into convergent hubs or polarized clusters. We theorize the dynamic interplay between structural expansion and cultural contraction and explore how this introduces an essential tension between the enjoyment and protection of difference.

**Society & Culture**

One of the most stable observations of the social and cultural world is that people linked together by social connections also tend to share cultural traits and attributes. When viewed as stable, global regions with shared language over the long durée, the identity of society and culture seem at best obvious, and at worst, a matter of definition. Untangling the recursive force between dynamic patterns of social awareness, communication and relations, on the one hand, and cultural tastes, expectations and ideas, on the other, has stimulated extensive theory, observation and experiment in anthropology, sociology, psychology, communication and related fields. Shared cultural forms have long been shown to drive the personal selection of new social network ties. The aphorism 'birds of a feather flock together' (McPherson, Smith-Lovin, and Cook 2001) suggests a widespread preference, termed homophily, for associating with persons like oneself (Lewis and Kaufman 2018). Social actors that possess the same preferences also consummate network ties with similar others as they naturally mingle in the same baseball stadiums, libraries, churches and chat rooms (Feld 1981).

Despite these powerful forces of social selection, another research tradition ranging from the social psychology of peer pressure to the sociology and statistics of collective intelligence has firmly demonstrated the recursive influence of social connection on shared cultural forms including preferences, ideas, and behaviors. Classic and recent studies of conformity have shown that social actors of low status often explicitly mimic the opinions and behaviors of social elites in order to fit in (Asch 1956; Danescu-Niculescu-Mizil et al. 2013), just as 21st Century computational social science studies of communication document linguistic copying to ingratiate and persuade (Danescu-Niculescu-Mizil et al. 2012; McFarland, Jurafsky, and Rawlings 2013). Even when cultural mimicry is unintended, social ties unleash communication and catalyze shared awareness that engage actors with one another's culture. In order to communicate, for example, conversationalists must adopt a common language or symbol system like Creole, Twitter-English or the proverbial "Bird"[1] and also a shared focus of communication (e.g., why did your car just cut me off?), which together collapse the space of ideas that can be imagined, considered and shared (Whorf and Chase 1956). Furthermore, social interactions can themselves generate novel situations, like romantic encounters or violent conflagrations, which activate cultural meanings and reinforce the conversations that gave rise to them. Cultural forms not only inspire the creation of new relationships, but new relationships can constrain, correlate and collapse the population of cultural forms.

In this paper, we do not attempt to solve the great causal riddle of society and culture, but rather to investigate, represent and interrogate a critical entailment of social networks' influence on cultural forms. Analysts have studied the process by which novel, bridging social ties can dislodge entrenched ideas and facilitate individual creativity and systemic innovation (Burt 2004). But bridging structural holes not only allows cultural forms to mix and react like chemicals poured into a common beaker. Increased social connectivity also forces cultural forms to compete for collective attention, and ultimately *contract via extinction* like animals and plants introduced into a common ecology via migration, or local languages introduced into a common global market via conquest and trade. Increases in cultural contact accelerate both cultural bricolage and cultural extinction. Here we focus on the process by which social connections collapse the population of cultural forms and its implications for theories of social integration and change.

---

[1] The raising of the middle finger as a sign of aggressive insult.

We first review recent studies that explore the implications of social connection on cultural, epistemological and ideological contraction in observational and experimental settings. Next we formalize these within the language of information theory. We then introduce a new geometric approach to data analysis in order to systematically examine associations between the social connectivity and the cultural diversity it can sustain in context of the epistemic culture of $21^{st}$ Century physics. We link networks of collaborative social relationships and cultural concepts through manifold learning into comparable, two-dimensional hyperbolic spaces or Poincaré disks. A Poincaré disk can render a hierarchical tree without distortion, allowing such spaces to reveal both the hierarchy and diversity of associations and expressions. Using metrics defined on these spaces, we empirically explore the association between shifts in the density of collaboration and changes in the diversity of topics discussed and ideas investigated. We find that collaboration drives new topical investigations, despite extreme long-term career commitments academic physicists must make in order to cultivate reputations and achieve promotion, acclaim and advance. Finally, we theorize the dynamic interplay between social expansion and cultural contraction and explore its implications for multicultural societies and the maintenance of diversity.

**Social Contact and Cultural Collapse**

A number of recent studies have experimentally and observationally identified the influence of network structure on the diversity of ideas. One large-scale laboratory experiment by Centola and Baronchelli explored the conditions under which cultural conventions spontaneously emerge and diversity collapses (2015). Participants were randomly assigned positions in a large social network. At each stage of the multistage experiment, two network neighbors were chosen at random to simultaneously assign names to a human avatar. They blindly attempted to coordinate agreement for reward in a real-time exchange of naming choices. Even though subjects had no knowledge about the population or awareness that they were engaging in global coordination, increases in network size and connectivity led to the emergence of a single winner as alternatives were driven to extinction. Homogeneously mixing populations experienced the most rapid evolution, emergence of agreement, and drop in diversity.

Research on the wisdom of crowds has explored the relationship between connectivity and collective wisdom. Crowds appear wiser than individual participants in stock markets, political elections, and quiz shows (Surowiecki 2004), but one recent experimental study demonstrated how social connections and the communication they unleash undermine crowd wisdom by reducing independence (Lorenz et al. 2011). In the experiment, subjects answered factual questions independently, and then revised their answers based on information about others' responses. The initially independent subject responses were accurate, but exposure to others' estimates collapsed opinion diversity and boosted individual confidence as subject's estimates converged with the crowd and away from the truth. Another experiment replicates this finding for subjects connected through both centralized and decentralized networks and involved in a complex estimation task (Becker, Brackbill, and Centola 2017). Centralized networks decreased the diversity of crowd opinions away from the truth, and decentralized networks decreased that diversity toward it, as the most accurate estimators were least sensitive to social influence.

In our own work, we have examined these patterns naturalistically in the large, epistemic culture of biomedicine (Cetina 2009). We extracted 51,292 published claims about drug-gene interactions from thousands of papers in the biomedical literature, then aligned them with a massive, robotically-assisted "high-throughput" experiment[2] to evaluate the likelihood of replication (Danchev, Rzhetsky, and Evans 2018). We deemed scientists that published papers on each drug-gene claim a scientific community, then extracted the direction of the drug-gene effect from each mention, along with prior literature the paper cited and methods it employed. We found that centralized communities, in which a few scientists collaborated on several papers sharing a common claim, were much more likely to draw upon the same literature, use the same methods, and claim drug-gene relationships that agreed with their own prior work. Moreover, claims generated by centralized communities were much less likely to replicate, because their experiments were not independent. Drawing on the same prior literature, they shared the same culture of expectations about what to expect from the experiment. Using the same techniques, they performed the same experiment repeatedly. Finally, with the same personnel—postdocs or principle investigators—they analyzed and interpreted the data in the same way. As a result, their findings were much less likely to be confirmed by a future experiment, just as they would be unlikely to manifest efficacy in the clinic. As with the experiments described above, more dense network ties led to less diverse experimental designs and less robust, replicable knowledge.

---

[2] NIH-funded LINCS L1000 experiment, which tested approximately 1.4 million drug-gene associations.

In related work, we have explored the degree to which scientists cluster, and the diversity of their interests contracts, as they select problems to research in science and scholarship (Rzhetsky et al. 2015). Drawing on millions of papers and patents published over 30 years, we estimated a model that identified the typical strategy for research selection in biomedicine, which considered the degree to which scientists respond to the popularity of topics researched by other. We showed that with increasing growth and density in the field, scientists and clinicians adopted more conservative research choices that clustered around more popular therapeutics, diagnostics and biological mechanisms with lower rates of publishable discovery. This observed strategy turned out to be efficient for initial exploration of the area because the hypothesis that successful drugs and diagnostics was more fruitful in the early stages of a field when little had been tested. This hypothesis led to pathological crowding in mature fields when much had been tested, found false, and yet remained unpublished due to the 'file drawer' problem (Rosenthal 1979; Scargle 1999), only to be re-investigated again and again. Nevertheless, the likelihood that a popular drug, diagnostic or mechanism would be useful in a new context was greater than that a rare or neglected one would be. In this way, conservative strategies support risk-averse scientific careers that require steady output, but are inefficient for advancing science as a whole. Through supercomputer experiments on a large sample of the network, we simulated thousands of alternative strategies that would have been much more efficient to explore mature fields, but these would have involved increased risk-taking for scientists. Although we often think of scientists as pursuing discoveries and innovation, this and related work reveals the growing role of tradition that emerges in science, as in other cultures, as it becomes a large, networked community (Foster, Rzhetsky, and Evans 2015; Uzzi et al. 2013).

The relationship between mutual awareness and the collapse of ideas was recently explored in a field experiment where 733 programmers competed in an algorithm development task under two alternative communication and disclosure regimes (Boudreau and Lakhani 2015). In the "open" regime, programmers shared the intermediate success of their code in a task involving the comparison and annotation of many genomic sequences for Harvard Medical School scientists. Intermediate disclosures efficiently limited experimentation and narrowed technological search, as teams crowded to reuse the most successful intermediate approaches. Under the closed or competitive regime, programmers competed without communication and a much more diverse range of approaches were cultivated. The problem was relatively simple, and so the open approach was more successful, but complex problems may yield the opposite result, as the linkage between early and ultimate success weakens. This experiment can be understood as an extreme comparison of two teaming regimes, where the "open condition" placed all developers into one, massively connected team, and the competitive condition kept them all apart, as singletons. Another recent study of our own examines the relationship between dense social connection and cultural collapse across technology, software and science. Large creative teams generate fewer cultural products per member than small ones (Shi, Foster, and Evans 2015). Nevertheless, analyzing teamwork from more than 50 million papers, patents, and software products, 1954-2014, we find that smaller teams engage in a much broader search than larger teams, on average, drawing upon older and less popular prior work as source material, and creating work that is much more disruptive to the status quo (Wu, Wang, and Evans 2017). By contrast, large teams converged to only the most recent, popular ideas, products and software. Attention to their own work, as a result, also tended to occur immediately. Together, this suggests that greater networking through collaboration in large, overlapping teams tends to collapse the diversity of the ideas and products explored.

This focus on system-level diversity contrasts but does not conflict with the large number of network analyses that focus on social connections for the diversity, creativity and ultimate success of the person or organization possessing them. In Granovetter's classic article "The Strength of Weak Ties" (1973) and his subsequent book, *Finding a Job* (1974), he demonstrated how broad, weak-tie networks enable the flow of novel employment information. Similarly, Burt has shown how managers with ties that bridge distinct social clusters within a company tend to generate broader and better strategic ideas (2004), and Stark and Vedres have shown how overlapping entrepreneurial ventures reap performance benefits (2010). In a related study of video game design teams, connections were particularly valuable for the creation of critically acclaimed and hit games, especially when they bridged cognitively diverse groups (de Vaan, Vedres, and Stark 2015). Another project reveals how articles that mix journals in scholarly production to simultaneously build on tradition and bridge distinct social contexts of research are rewarded with outsized citations (Uzzi et al. 2013). At a higher level of analysis, research on innovation within cities has demonstrated a superlinear scaling of innovative activity (Bettencourt et al. 2007; Arbesman, Kleinberg, and Strogatz 2009) and diverse information intake (Schläpfer et al. 2014) with city size around the world.[3] What remains unevaluated in these studies is how the outsized success of certain creative or successful combinations influence future creativity by driving attention away from alternative and isolated cultural forms (Evans 2008). Moreover, while studies have shown that the information benefits from bridging structural divides would erode if everyone pursued a bridging strategy (Buskens and van de Rijt 2008), the perceived ubiquity of social and cultural barriers has prompted many to advocate greater interdisciplinarity and boundary-crossing creativity (Bauer 1990; Braun and Schubert 2003) without considering its cost for the cultural diversity that might limit returns from future expeditions. Most work on connection and creativity focuses on current, individual-level advantages, rather than counting costs for the entire system over longer time scales.

---

[3] Research on cities often focuses on not only the combination of difference, but its creation through specialization, but the equation between cultural difference generation and erosion is never explicated.

Some scholars have begun to consider the limits of systemic creativity that arise from ubiquitous and repeated connections, but typically at local rather than global scales. Studies of creative teams reveal the diminishing marginal creativity of repeated collaborations in scientific and scholarly research, product innovation (Guimerà et al. 2005), and Broadway musical creation (Uzzi and Spiro 2005). March's simulation of divergent organization learning styles is an exception (1991), which does consider global consequences of communication by contrasting short-term exploitation of sure bet opportunities with long-term, global exploration of new, uncertain ones. Here we build on March's work by theoretically articulating and empirically illustrating how excessive communication can simultaneously facilitate the exploitation of short-term learning opportunities, and limit global exploration and the possibilities of future discovery. Across cultural, artistic, scientific and technological domains, we see that greater social networking and connectivity leads to decreased diversity in the population of cultural forms. This suggests that as communities enlarge and their social connections intensify, their collective carrying capacity for cultural diversity shrinks. Most of the studies described above, however, do not represent social actors and cultural forms in a common framework to enable direct measurement of this relationship.

**The Information Theory of Cultural Collapse**

Shannon's information theory furnishes a formal model of communication and information (Shannon and Weaver 1963) that illuminates the theoretical and observed relationship between social connection and cultural contraction. In information theory, we can associate communication with social connections, as pre-existing "information channels" are required to exchange messages. Information, in Shannon's theory, is a message passed between sender ($X$) and receiver ($Y$) that answers a question or otherwise resolves uncertainty for the receiver. If we define each cultural form as one of many possible outcomes from communication, then these messages can be understand as the expression and transmission of tastes, expectations, ideas, behaviors and artifacts. Information $I$ is equivalent to the "surprisal" or surprise of experiencing a given message or cultural communication, such that an improbable or unexpected message has more information than an anticipated one and is experienced by the receiver with greater surprise. Information, then, is the news quality of a message. If senders and receivers of information mingle through repeated back-and-forth communication, however, the *mutual information I(X;Y)* of their messages—the information common to both *X* and *Y*, modeled as two random variables—increases. This increase in mutual information necessitates a drop in *joint entropy H(X,Y)*—the average information produced by both actors communicating together (see Figure 1, top panel). This relationship assumes that individuals exhibit a conserved number of cultural forms and that what forms they possess, they express.

______________________________

Figure 1 about here

______________________________

The pattern for two communicating actors generalizes to an arbitrary numbers of communicators linked via social network. In larger networks, the inverse relationship holds between the multivariate mutual information and multivariate joint entropy, with some slight technical amendment. The multivariate mutual information can be positive or negative, and so for the identity to remain precisely equivalent in a larger network and that of a single dyad, the joint entropy shrinks linearly with growth of the union between (1) the multivariate mutual information (e.g., *I(X;Y;Z)* in a three-person network; see lower left panel in Figure 1), and (2) all conditional mutual information measures (e.g., *I(X;Y|Z)*, *I(X;Z|Y)*, *I(Y;Z|X)* in a three-person network; see lower left panel in Figure 1). This reveals that when increased channels of communication are unleashed by a denser pattern of social relationships, the cultural information passed between them shrinks in direct proportion to their shared or mutual information.

This information theoretic rendering of the relationship between structure and culture reinforces the mutually constitutive relationship between individual-level innovation unleashed by novel social connections and communication (Burt 2004) and the system-level diversity depressed by it (March 1991). Connecting disconnected societies enables cultural mixing in the short-term, but through competition and selective cultural extinction, future periods have fewer cultural forms to mix. The more socially and culturally distant the groups connected, the greater the immediate innovation potential for the socio-cultural "arbitrageurs" that connect them, but the larger decrease in system-level diversity that results from putting more cultural forms in contact and competition. This suggests that the second law of thermodynamics has a manifestation in culture, such that the distribution of cultural forms in a social system becomes more *even* as social relationships increase the mutual exchange of information. Next we explore geometric representations of social and semantic structures to reformulate our thesis regarding social contact and cultural collapse.

**The Hyperbolic Geometry of Social and Semantic Networks**

Social network analysts have historically considered only the topology of social networks, and not their intrinsic geometry. At the limit, a graph with *n* nodes would require up to *n*-1 Euclidean dimensions to represent geometrically, and so geometric embedding was not perceived as reducing the complexity of network data. The Johnson-Lindendstrauss theorem (1984), however, proved that networks can be represented with minimal distortion in far fewer (~log *n*) dimensions. The rise of autoencoders, or neural network models that automatically learn to embed data from one representation into another, alongside improvements in high performance computing have recently enabled analysts to efficiently and practically place social networks into much lower dimensional geometric representations (Nickel and Kiela 2017; Chamberlain, Clough, and Deisenroth 2017).

Language and meanings, on the other hand, have long been represented within continuous geometric spaces, initially with semantic surveys and factor analysis (Osgood, Suci, and Tannenbaum 1964). Later, Latent Semantic Indexing and Analysis used singular value decomposition (SVD) to reduce the dimensionality of word-document matrices and placed words in a space defined by their coarse semantic contexts (Furnas et al. 1988). More recently, efficient auto-encoders like Word2Vec (Mikolov, Sutskever, et al. 2013) and GLoVe (Jeffrey Pennington, Socher, and Manning 2014) have replaced SVD, enabling analysts to simultaneously use far larger text repositories with far smaller linguistic contexts than before. The resulting word embeddings reveal semantic compositionality (Bolukbasi et al. 2016) and recover widespread cultural associations and biases (Garg et al. 2018; Kozlowski, Taddy, and Evans 2018).

In order to directly test our hypothesis regarding the extension of social networks and the collapse of cultural forms, we propose embedding them within the same geometric space using autoencoders. This will allow us to evaluate how the extent of social connections influences the size and structure of cultural forms. Nevertheless, alternative geometries have been use to represent social and cultural networks. Here we review alternatives and argue that hyperbolic geometry allows us to most naturally and efficiently represent social and semantic space.

**Hyperbolic Geometry and Structural Representation.** Statistical physicists borrowed and repurposed hyperbolic space from its first major scientific use to represent the expanding physical universe, in order to consistently capture the network topology of complex social and cultural systems (Papadopoulos et al. 2012; Krioukov et al. 2010). Euclidean geometry was constructed to obey Euclid's fifth postulate, which states that within a two-dimensional plane, for any given line $\ell$ and point p not on $\ell$, there exists exactly one line through p that does not intersect $\ell$. The result is a space dimensionalized by "straight" vectors and "flat" planes having zero curvature, such that a drawn triangle's angles add up to 180° and the Pythagorean theorem holds.[4] Curved geometries violate Euclid's fifth postulate, which is not derivable from other postulates. For example, in a 2-dimensional elliptical geometry, which has positive curvature and can describe the surface of the earth, all lines through p intersect ℓ, and any two lines perpendicular to ℓ intersect at a "pole". The sum of a triangle's interior angles on an elliptical surface is always greater than 180°. Consider a triangle etched onto earth's surface with one side stretching along a quarter of the equator and the other two sides meeting at the South Pole. All three angles will equal approximately 90°, summing to 270°.

Hyperbolic space, by contrast, has negative curvature, such that infinitely many lines may go through *p* without intersecting ℓ. A saddle, mountain pass, coral reef, or crocheted frill furnish examples of negatively curved surfaces. A triangle etched upon them will have angles that sum to less than 180°. As two intersecting lines diverge at a constant rate in Euclidean geometry (e.g., the Pythagorean theorem), and converge in elliptical geometry, they diverge exponentially in hyperbolic geometry such that there is "more space" in a 2-dimensional hyperbolic disc than a Euclidean circle or elliptical globe. In Euclidean geometry, the circumference of a circle with radius *r* equals *2πr*, but in hyperbolic geometry it is always more and elliptical geometry always less. Figure 2, panel A presents M.C. Escher's print *Circle Limit IV* and panel B shares the underlying Poincaré disk model, the projection of a 2-dimensional hyperbolic space onto a 2-dimensional Euclidean space, cited and recreated from (Dunham and Others 2009). Escher's print alternatively tessellates angels and devils on the disk, *ad infinitum*, and reveals how the space becomes exponentially more dense as the radius increases.

Nicholas Lobachevsky, one of hyperbolic geometry's 19[th] Century discoverers, argued that the universe might not be flat, but rather hyperbolic (Bonola 1955), and Einstein's general relativity intensified interest in the curvature of the universe by positing that mass and energy bend spacetime. In general relativity, mass and energy can be used to determine the universe's curvature (Ω) as the average density of the universe divided by the mass energy required for it to be flat. Despite recent estimates that put Ω at approximately 0 (Biron 2015), spacetime in the vicinity of massive bodies such as stars, solar systems and galaxies is positively curved, and space between massive bodies, in dark energy dominated areas[5], is negatively curved (Krioukov et al. 2012).

Statistical physicists, computer and network scientists have recently applied hyperbolic geometry as a compact representation of social networks (Papadopoulos et al. 2012; Krioukov et al. 2010) and semantic hierarchies (Nickel and

---

[4] The Pythagorean theorem states that the square of the hypotenuse of a right triangle (the side opposite the right angle) is equal to the sum of the squares of the other two sides.

[5] As dark energy is constant over the universe, regions with little mass are proportionally "dominated" by dark energy.

Kiela 2017; Chamberlain, Clough, and Deisenroth 2017). A Poincaré disk can represent a branching tree structure without distortion, and higher dimensional hyperbolic spaces can perfectly capture hierarchies with multiple heritage, such as directed acyclic graphs. As a result, negative curvature and diverging, branching structure of hyperbolic space enables us to model the hierarchy and sparse, bridging structures common to many realistic social and semantic networks in far fewer dimensions than a comparable number of Euclidean dimensions (Nickel and Kiela 2017).

**The Hyperbolic Geometry of Social Structure**. Euclidean space is constrained by the triangle inequality such that for any three sides of a triangle, $a + b > c$. This representational feature forces transitivity, such that friends of friends must be friends. Despite the resonance of this principle with social psychological theories of intimate networks like Heider's balance theory (Heider 2013), this property makes it difficult to represent large scale social networks that manifest substantial hierarchy and intransitivity (Feld 1981). All social network theories that focus on the importance of bridging ties, including canonical "strength of weak ties" (M. S. Granovetter 1973) and "structural holes" (Burt 2004, 2009) arguments, highlight the importance of violating transitivity for the flow of information and advantage. Social network actors with higher betweenness centrality (Freeman 1977) uniquely own more ties and bridge more structural holes. They play the role of knowledge brokers by connecting people disconnected from each other. In social systems, broken triads accumulate into the hierarchy of power structures, which allow for the division and aggregation of labor, scaling-up institutions to coordinate complex social action. Hyperbolic geometry, assuming constant, negative curvature and allowing triangle inequality is a strong candidate for modeling these asymmetric structures. Because of the expanded area or "ruffled edge" of hyperbolic space, two nodes can simultaneously be close to the same central, bridging node and far from one another.

Statistical physicists first proposed the use of a Poincaré disk to embed complex social networks (Papadopoulos et al. 2012; Krioukov et al. 2010). On the disk, the position of each node $i$ is defined by two parameters in Euclidean polar coordinates, radius and angle $\theta$. The radius quantifies position in the hierarchy: nodes of small radius literally hold a central position in the hierarchy. The angle between two nodes $\theta -$ quantifies their proximity or structural similarity.

The initial algorithm used to represent social networks in hyperbolic space was slow and inefficient (Papadopoulos, Psomas, and Krioukov 2015). The primary bottleneck involved the Hamiltonian Monte Carlo method used to estimate node positions, which repeatedly calculated the gradient or fit between network and geometric embedding and became burdensome as datasets scaled up. This changed dramatically when computer scientists discovered that artificial neural networks were powerful tools for embedding optimization. Moreover, analysts have demonstrated that low dimensional hyperbolic embedding improve upon higher dimensional Euclidean embedding for many tasks, including predicting the collaboration between scientists in weighted undirected collaboration networks (Nickel and Kiela 2017; Chamberlain, Clough, and Deisenroth 2017).

Hyperbolic space enables the simultaneous modeling of network clustering and hierarchy. The small-world property suggests that many real-world networks are locally dense, but with global reach. Local density is characterized by a high clustering coefficient, such that the friend-of-a-friend is also a friend (Watts and Strogatz 1998). Global reach is measured by a small network diameter, or the average shortest path between all node pairs. Many social networks reflect this property, such that diameter increases only linearly as number of nodes increase exponentially. The "small-world model" begins with a locally dense, globally disconnected network and rewires it by connecting randomly selected pairs (Watts and Strogatz 1998). Only a few of these rewired connections, empirically mirrored in migration and global communication, are required to dramatically decrease the world's diameter. This random-rewiring process introduces intransitivity and increases network hierarchy, because long-distance connections are "weak ties" (M. S. Granovetter 1973) and the social actors who disproportionately create them will have high betweenness as their friends typically remain disconnected.

Another important property of empirical networks is their "scale-free" diversity of connections, a property that traces a specific form of network hierarchy. The distribution of connections in most realistic social networks has a long-tail, such that a few social actors own most connections, while most actors possess only a few. The "scale-free" property implies that when scaled by the correct exponent $\tau$, the logged distribution of social connections can be represented as a straight line, such that any social actor in the network has the same relative proportion of those with immediately more and less connections. If $\tau = -1$, then if one person had a million Twitter followers, a million people would have one follower. Following Herbert Simon (Simon 1955), Barbási and Albert proposed a network mechanism they termed "preferential attachment" to explain the long-tail distribution of social connections (Barabasi and Albert 1999). In this model, network members prefer attachment to popular actors. As consequence, initially random degree differences between actors become amplified across periods so that random networks become "scale-free", dominated by high degree actors. This model predicts increases in network hierarchy, as high degree nodes are also high in betweenness. Low degree nodes connect to high degree nodes but not each other. As such, the preferential attachment model only generates graphs that are globally connected, but not locally dense.

For a decade after publication of "small world" and "preferential attachment" network models (Watts and Strogatz 1998; Barabasi and Albert 1999), network scholars sought to produce a model that reproduces both "small-world" and "scale-free" patterns (Song, Havlin, and Makse 2005; Zhang et al. 2015; Barabási 2009). In 2012, Papadopoulos and colleagues suggested that the key to reproduce both patterns is to translate the complexity of linking dynamics into their underlying geometry (Papadopoulos et al. 2012; Krioukov et al. 2010). They demonstrated how growing networks in a hyperbolic space with trivial linking dynamics (nodes are connected to each other within radius *r*) will exhibit both "small-world" and "scale-free" properties. This approach exploited the asymmetry of hyperbolic space such that an actor at the center can reach all other nodes but actors at the periphery are exponentially distant from one another, and must often go through those at center to reach others. Krioukov et al concluded that the hyperbolic space was characterized by two directions, the "similarity" directions along the periphery, along which nodes are locally clustered, and the "popularity" direction pointing from center to periphery.

Building on this work, we argue that the second direction should be understood as social "hierarchy" rather than "popularity", as it models intransitivity by exploiting the asymmetric nature of hyperbolic space. Central nodes in hyperbolic space bridge peripheral nodes with high betweenness. Hyperbolic spaces can be viewed as a continuous branching tree, in which the central network node corresponds to the root whose critical feature is high betweenness and not high degree. Our claim is supported by Chamberlain and colleagues who embedded social networks into hyperbolic spaces and found that high-betweenness nodes became central, whereas high degree nodes with low betweenness were pushed to the periphery (Chamberlain, Clough, and Deisenroth 2017).

**The Hyperbolic Geometry of Language and Culture.** The distributional hypothesis in linguistics claims that words used in the same contexts tend to purport similar meanings (Harris 1954), a concept popularized by Firth's gloss that "a word is characterized by the company it keeps" (1957). As one of the most important principles in computational linguistics, the cognitive origins of the distributional hypothesis have been confirmed by psychological experiments (McDonald and Ramscar 2001). The distributional hypothesis has found wide application in natural language processing, from topic modeling to machine translation (Blei, Ng, and Jordan 2003). This principle structurally corresponds to the importance of local clustering in social networks (Lorrain and White 1977; Newman 2010; Salton 1989).

The distributional principle is also the core principle of word2vec, a neural network model trained to predict word co-occurrence for rich semantic representations in a multi-dimensional Euclidean space (Mikolov, Chen, et al. 2013). Word2vec operationalizes the distributional hypothesis by representing words and contexts in a vector space, using neural networks as an optimization model for vector estimation. *Word2vec* can operate under two distinct model architectures: continuous bag-of-words (CBOW) or skip-gram. Under the CBOW architecture, the corpus is read line-by-line in a sliding window of *k* words, such that for each word in the corpus, the algorithm aims to maximize classification of the center word *n*, given the surrounding *k* words. The skip-gram architecture works similarly, except that instead of predicting a word with context, the skip-gram architecture predicts context given word. Related embedding approaches take into account additional information such as the "global" proximity of words within an overarching document (J. Pennington, Socher, and Manning 2014) or subword letter sequences in surrounding words (Joulin et al. 2016). By representing words using context word distributions, word2vec quantifies semantic similarity by generating similar vector space coordinates for semantically related words.

Word2vec represents words in *k*-dimensional vectors, or the weights of all *k* neurons in the neural network's hidden layer, where *k* is typically 200 to 300 dimensions. Word embedding models are sometimes considered "low dimension" techniques relative to the number of words used in text (e.g., 20,000) because they reduce this very high dimensional word space. Word2vec reduces data complexity from $n \times m$ in a corpus with vocabulary length *n* and context length *m*, to $n \times k$, where $k \ll m$. By using word co-occurrence as the basis for distance between words in a *k*-dimensional Euclidean space, word2vec minimizes the dot product and angle between the vectors of co-occurring word pairs. In comparison with one, two or three dimensional models common in representing culture (Bourdieu 2013), however, these spaces are sometimes considered "high dimensional" in that they produce much more complex and accurate associations, as demonstrated by many recent analyses (Kozlowski, Taddy, and Evans 2018; Garg et al. 2018; Bolukbasi et al. 2016; Caliskan, Bryson, and Narayanan 2017).

The resulting spaces allow the analyst to engage in cultural reasoning with vector operations, such as $\vec{king} - \vec{man} + \vec{woman} \approx \vec{queen}$ or $\vec{Vietnam} + \vec{capital} \approx \vec{Hanoi}$. This property, termed additive compositionality (Mikolov, Sutskever, et al. 2013) or meaning calculation (Tian, Okazaki, and Inui 2017), hinges on the Euclidean geometry underlying word2vec.[6] In harmony with the computational theory of mind (Piccinini and Bahar 2013), vector addition acts like an AND logical operator, selecting common context words, while subtraction acts like a NOT operator, selecting context words distinct from the word subtracted. In this way, word2vec and related models represent language to reveal cultural meaning through position and proximity.

---

[6] The softmax filter often used to estimate word2vec models imposes an exponential function that transforms the addition of word vectors into their pointwise multiplication, and the subtraction of word vectors into their pointwise division.

Clustering and hierarchy observed in social networks are also present in language (Liljeros et al. 2001; i Cancho and Solé 2001). Because word2vec translates co-occurrence frequencies into angles between word vectors, if the word *drive* co-occurs frequently with words *car* and *disk*, the measured angles $\angle \vec{car}\vec{drive}$ and $\angle \vec{disk}\vec{drive}$ will both be small. Because Euclidean space enforces the triangle inequality, the model can only allow for $\angle \vec{car}\vec{disk} < \angle \vec{car}\vec{drive} + \angle \vec{drive}\vec{disk}$. This proposes a modest semantic similarity between *car* and *disk*, despite their infrequent empirical co-occurrence. More importantly, intransitivities may accumulate recursively. *Drive* may be the bridge between *car* and *disk*, but there other "parent nodes" like *vehicle* could bridge *drive* with other, dissimilar words, such as *skateboard* and *zeppelin*. We argue that this accumulation of intransitivities builds up hierarchies of meaning that constitute a "ladder of abstraction" (Hayakawa and Hayakawa 1990). Open, nontransitive triplets represent building blocks in this hierarchy, which enables human thinkers to simultaneously differentiate meanings within small, context-dependent sets (MILLER and A 1956) and still move up and down the ladder of abstraction to understand and generate complex meanings.

Embedding corpora in a hyperbolic geometry could facilitate the modeling of semantic hierarchy as well as clustering, which word2vec and its cousins already perform well. Embedding language networks and hierarchies, like the digitally networked English dictionary WordNet, within hyperbolic geometry requires fewer dimensions than in Euclidean geometry to infer the same complex relationships between words (Fellbaum 2005; Chamberlain, Clough, and Deisenroth 2017; Nickel and Kiela 2017) because the space better reflects the intrinsic geometry of the data, and also because there is "more space" in the hyperbolic representation.

For these reasons, we embed social and cultural relationships within comparable hyperbolic spaces. This will allow us to directly interrogate the relationship between patterns of social connection and cultural diversity. We analyze these associations within the epistemic culture of academic physics, as represented by the complete 21st Century publications of the American Physical Society (APS), a society organized at Columbia University in 1899 "to advance and diffuse the knowledge of physics". Insofar as the hyperbolic embedding of socio-cultural structure represents a kind of *physics of sociology*, we felt it might be appropriate to recursively use it to perform a *sociology of physics*. We specifically examine patterns of collaboration between universities in terms of their publishing physicists, and the diversity of physics topics those universities publish. We use published physics rather than a more ephemeral socio-cultural context like chit-chatters on Tumblr, Twitter or Facebook largely because it is much more difficult to identify individual persons and trace them persistently in social media settings, even though it may be easier to follow micro-social and cultural interactions in those settings. Because academic physicists have a strong incentive to build persistent reputations through publishing within a single area, our data represents a very conservative case from which to examine the impact of social relationships on cultural forms. If we observe that relationships tend to drive topic selection and contraction among academic physicists, it is likely to be amplified elsewhere.

**Data**

APS data publication is a freely accessible online dataset (https://journals.aps.org/datasets) that contains the meta-data for articles published in *Physical Review Letters*, *Physical Review* (*A, B, C, D, E and X*), and *Review of Modern Physics*, some of which predate the APS, going back to 1893. In order to analyze collaborations between scholars and institutions, we first needed to disambiguate author and institution names: the same name may refer to multiple people, and distinct affiliation names reference the same institution. To extract unique author names, we matched the APS meta-data with Web of Science (WOS) data produced by Clarivate Analytics, using the DOIs (Digital Object Identifier) of papers, 2001-2013. In WOS, we used a hybrid algorithm that exploits both metadata and citations to disambiguate authors and obtain 10,051,491 scholars who contributed to 22,177,224 papers in 115 years (1900-2014). We used the 2017 Open Researcher and Contributor ID (ORCID) dataset to validate our results and find a high precision (78%) and recall (86%) among the 118,094 ORCID scholars of three or more self-identified papers. From this disambiguated data, we successfully identified 372,495 authors from the analyzed 359,395 APS papers.

To resolve institution name ambiguities, we extracted affiliation names and fed them into the Google Geolocation API to identify whether the same latitudes and longitudes are returned. Google Geolocation provides results within 25 feet if the address translation is correct. Leveraging this approach, we merged 330,560 institution names into 55,317 unique ones. Finally we connected name-disambiguated institutions to name-disambiguated authors in each year, enabling authors to change institutions across years. In total, we obtain 37,2495 authors from 28,754 institutions. The most institutionally connected author has 26 institutions, and 83% of authors have only a single one.

We weighted author collaboration networks by iterating over all papers for a given year and connecting scholars that collaborated on the same paper. In this way we obtained ten networks of author collaboration for 2002 through 2011. Next, we collapsed the networks by merging scholars from the same institutions into single nodes and then aggregated author collaboration edges to connect institutions. Edges between scholars from the same institution are retained as self-loops. To compare institution networks across years, we fixed the nodes by selecting the top 300 institutions, but 10 were removed due to missing data and so we analyzed collaborations between the most prolific 290 institutions. The number of edges increased from 3,331 (202) to 7,647 (2011), and the total weight increased from 29,117 to 94,336 over this period. After we constructed the ten yearly institution collaboration networks, we embedded each into a 2-dimensional hyperbolic space using the Poincaré disk model (Nickel and Kiela 2017). As anticipated, we found that the embedding space captured both the hierarchy among institutions and the similarity between them, as shown in Figure 1.

The Physics and Astronomy Classification Scheme (PACS) was developed by the American Institute of Physics (AIP) and has been used in the Physical Review journals since 1975 to identify fields and subfields of physics. PACS is a hierarchical partitioning of the full spectrum of topics in physics, astronomy, and related sciences. Each PACS code is a six-digit number, in which the first digit indicates one of ten subfields to which it belongs (see Figure 2). We constructed the PACS code co-occurrence network by iterating over all 359,395 papers in our dataset and connecting PACS codes that appeared together in the same paper. The network contained 5,856 nodes, 244,106 edges, with a total edge weight of 1,124,724. After we constructed this PACS code co-occurrence network, we embedded it into hyperbolic space using the Poincaré disk model, comparable with the institutional collaboration embedding described above. As with the organization network, we found that the representations of PACS codes in a 2D hyperbolic embedding captures both their hierarchy and similarity, as shown in Figure 1.

**Methods**

We use the Poincaré Embeddings algorithm (https://github.com/facebookresearch/poincare-embeddings) developed by Facebook AI scientists Nickel and Kiela (2017) using PyTorch. This algorithm automatically learns hierarchical representations of nodes in networks by training on positive and negative samples by using a shallow, three-level neural network auto-encoder. We embedded our (1) university collaboration network across publications in all time periods (2002-2011), (2) separately for each year, and (3) the co-occurrence network of all PACS codes across publications in all years, using a 2-dimensional Poincaré disk, with a learning rate of 0.5 and a negative sample size of 50.[7] This produced hyperbolic embeddings in which each node $i$—a university in the collaboration embedding and a PACS code in the cultural embedding—has radius   and angle $\theta$ . Nodes of small radius literally hold central positions in the circularly arrayed hierarchy. Angles between two nodes, $\theta\ -\ $ , quantifies their social or cultural difference.

---

[7] We specified a batch size of 30, which limits the size of examples from data on which the algorithm trains. We trained the model on 300 sequential epochs, of which the first 20 were "burn-in" epochs used only to initialize the model. The model was evaluated every 5 epochs, and every time gets updated if it has lower loss against the validation data. The co-occurence of collaboration edges in our network are undirected and so we set the parameter *symmetrize* to be true. In addition, the *epsilon* parameter is set be no smaller than 1e-5 such that the maximum hyperbolic radius is 1-1e-5.

In order to calculate the position $\theta$ for each university $i$ in the cultural or PACS code space, we selected the peak or mode of the Gaussian kernel density estimation for all of its weighted PACS codes in a given year (see Figure A4) from the PACS code positions estimated from all years, 2002-2011 (see Figure 2D). To estimate each university's level in the topical hierarchy, we calculated the average of their ten smallest values —their extreme values. Due to asymmetry in the hyperbolic space, most PACS code values of approached 1 and so neither mean nor median well characterized the difference in cultural hierarchy between universities and so we evaluated difference in their extremes. With this estimation, each university $i$ has a unique $r_i$ and $\theta_i$ in both collaboration and topical space for each year.

These preliminaries set us up to directly explore the relationship between social and cultural—collaboration and subject focus—in physics. We first did this by taking each pair of universities and calculating their social and cultural hyperbolic distances at every time point, which generated two time series. We then correlated and evaluated these time series pairs with Granger causality regression with one-year lags in order to explore the complex relationship between social and cultural factors. This suggests that when two institutions increase their social contact, they contract towards one another. We also performed the same analysis between each university and all other universities in the system at every time point to explore the connection between social relationships and the contraction of the entire cultural system. In the appendix, we validate our findings with traditional network measures of social density and information theoretic measures of cultural diversity.

**Findings**

**The Social and Cultural Geometry of Physics.** The hyperbolic space can be viewed as a mountain with an exponentially increasing base. The center corresponds to the mountain peak. The closer to the peak, the more easily one can view others at the base, who may not be able to see, reach or communicate with one another. By embedding both institutional and PACS networks into comparable hyperbolic spaces, we make visible the duality between social connectivity and cultural diversity, and enable evaluation of the hypothesis that institutions moving together in the social space will converge to more similar topics in cultural space (Figure 2).

---

Figure 2 about here

---

Figure 2, panel C shows the hyperbolic embedding of the institutional collaboration network in 2011 using the the Poincaré disk model. The 290 institutions displayed are colored by region. Panel D shows the hyperbolic embedding of the aggregated PACS code co-occurrence network using the the Poincaré disk model. The analyzed 5,819 PACS codes (dots) are colored by the 10 subfields to which they belong. We employed Gaussian kernel density estimation to visualize the concentration of the angles of PACS codes in each subfield (see Figure A4 for the details). To compare across subfields, we also rescaled the estimated distributions such that the area covered by the distribution curve is proportional to the total number of papers published by 290 institutions in each subfield from 2002-2011.

In order to summarize the global social and cultural structure of 21$^{st}$ Century physics, we calculated and graphed the radius and angle in embedded social space for each of the 290 top physics publishing universities and in cultural space for thousands of PACS codes. We found that in both social and cultural disks, the radius systematically quantifies the "hierarchy" of university physics departments in the inter-university collaboration network, and angles quantify the structural equivalence between nodes in the network. As demonstrated in Figure 2C, high-ranking institutions from various global regions occupy the central area in the social space, including Columbia University from North America, Peking University from Asia, and Australian National University from Australia. Geological groups emerge and cluster together on the periphery, including institutions from Canada (University of Waterloo, McMaster University, McGill University, and the University of Alberta), Japan (Kelo University, Waseda University, Osaka City University, and Tokyo Metropolitan University), Israel (Hebrew University and Hebrew University of Jerusalem), and Italy (the Universities of Catania and Radua).

In Figure 2D, most PACS codes group together on the perhiphy, close to other codes within their own and related subfields. A few are more general such that they are placed closer to the hyperbolic center. The most concentrated PACS codes include "The Physics of Elementary Particles and Fields" (red) and "Nuclear Physics" (orange), which nontrivially overlap in the lower left quadrant of the Poincaré disk. The tall peaks of density distribution centered on these fields suggest that their codes are nearly always used with others of the same field in research articles. They rarely combine with other field codes. Situated at the antipodes of the disk, in the upper right, are "Condensed Matter: Structural, Mechanical and Thermal Properties" (light blue) and "Condensed Matter: Electronic Structure, Electrical, Magnetic, and Optical Properties," which each have a higher $\theta$ and are substantially more spread. Nuclear physics and the physics of elementary particles focus on individual particles and their properties, including their decay. Condensed matter physics focuses on interactions between particles, often in the context of large, naturalistic materials. It may not do too much violence to the epistemic cultures of physics to say that particle physics is the *economics* of the physical sciences, flush with cash for large-scale experiments, focused on the essential monads of existence in artificial isolation and furnished with a sense of superiority that their work is more fundamental and important than their neighbors (Fourcade, Ollion, and Algan 2015).

Condensed matter physics might then be viewed as the *sociology* of the physical sciences, fixed on social interactions between particles in complex, overlapping physical fields, often embedded in the rich context of materials and phenomena that can be macroscopically experienced by ordinary humans, like sandpiles (Jaeger and Nagel 1992), coffee rings (Deegan et al. 1997) and the physics of food (Poon 2002). Consider the Ising model from statistical mechanics that sources from "soft matter" concerns and models the emergent property of phase transitions in ferromagnetic materials as atomic spins arranged in a graph. The graph arrangement allows each spin to interact with its neighbors. This model is frequently used by social network analysts and physicists to model human social interaction (Klemm et al. 2003; Wasserman and Pattison 1996). Insofar as physics increasingly tends to cleave between a focus on particles or interactions (see Figure 3B), we leave it to the reader to decide which represents the angel and which the devil in Figure 2A.

Plasma physics, fluid dynamics and classical mechanics lie off of the fundamental particle / soft matter axis, but have also been much less intensively published in 21$^{st}$ Century APS publications.[8] The most spread field is not a field at all, but rather "Interdisciplinary Physics and Related Areas of Science and Technology" (deep blue), which diffusely spreads around the entire system and whose codes are frequently connected to those from other fields. That these codes are not often close to the center of the disk suggests that rather than "Interdisciplinary", they are more likely "Multidisciplinary" codes that interact with other specific fields at the periphery, but do not integrate those fields into the broader system of physics. The field codes closest to the center are those from "General" (red), which is not a category at all, but rather was derived from the first two decimal places of each PACS code and so, by definition, connects different subfields of physics by forming the trunk of their hierarchical tree.

We map changes in the social hierarchy of university physics collaborations over time, from 2002 until 2011 in Figure 3A, where universities are colored by region. Here we see strong and consistent geographical clustering in collaboration over time, with North American, European, and Asian universities tending to collaborate with one another. Australian and South American Universities are dispersed as satellites of geographically diverse, dominant universities from Asia and the West. Within the hierarchy, North American institutions remain the most central collaborators throughout, with the lowest average radius *r*, but Asian universities experienced the greatest change in *r*, migrating rapidly from the periphery toward the collaboration network's center.

---

[8] Some fields like plasma physics have suffered in the U.S. and Europe from encumbrances for the international collaboration, ITER, the International Thermonuclear Experimental Reactor.

We also chart changes in the cultural position of physics universities over the same time period in Figure 3B, with universities colored by region, as in 3A, but plotted against the subfield locations represented in Figure 2D. It is interesting that the representative positions of most university physics departments cluster near the subfield of "Condensed Matter Physics" (45 degree) or "The Physics of Elementary Particles and Fields" (225 degree). Moreover, the field of university physics departments tends to collapse toward polarization between the physics of particles and of complex materials over time. In 2002, a scattering of European and a few U.S. universities have a wider mixture of PACS codes, focusing on plasma or fluid physics, but by 2011, they tend to represent some mixture of particle or systems physics. North American Universities appear most likely to specialize in particle or systems physics, while European and Asian Universities represent a much more even mix of the two polar branches.

**Social connection and cultural contraction.** When we compare the social and cultural hyperbolic distances between every pair of universities over all years, binned, we see a very strong positive association, as graphed in Figure 3C. This suggests that social and cultural distance are tightly bound together across all years. We also see that the range of distances is much smaller in the social than the cultural world. Nevertheless, the range of the two distances are very different. Topical distances vary between .4 and 1.4, while social distances move from .5 to 8. This difference reflects the deep topical inertia in academic physics, while patterns of collaboration can shift dramatically from year to year. We also see that the cultural worlds of university physics are dramatically contracting over time, nearly halving for the least connected institutions between 2002 and 2011.

---

Figure 3 about here

---

Next, we explore the association of collaboration and focus over time. For each pair of institutions, we estimated an OLS regression to identify the slope representing change in social distance over time. The coefficient for time, $\beta_{sij}$, should be interpreted as the speed of social convergence or divergence between any two institutions *i* and *j* across the 2002-2011 time series. Then we separately estimated the slope of cultural change over time, where the coefficient on time, $\beta_{cij}$, reflects the speed of cultural convergence or divergence between that pair of institutions. The Pearson correlation coefficient between the two diverging speeds, $\beta_{sij}$ and $\beta_{cij}$, is 0.98 (P-value < 0.001), providing strong evidence that in the context of academic physics, collaboration is associated with topical convergence (see Figure 3D). In summary, not only are university physics collaborations moving with their evolving interests and expertise, but they are moving the same way over time.

We also explored the relationship between the position of each institution $i$ within the hierarchy of institutions, $r_{si}$, and their position within the hierarchy of cultural commitments or physic topics, $r_{ci}$. Even though an institution's cultural or PACS code radius is only determined by the extreme values (see Methods), the inset of Figure 3C reveals a strong correlation between university positions within the two hierarchies. When we evaluate these dynamically as the relationship between the change in social radius and cultural radius over time, we find a Pearson correlation coefficient of 0.85 (P-value < 0.001).

Next, we explore the directional prediction of social and cultural distance, and their component parts—radius and angular distance—over time, by lagging each variable by one year and predicting the others. Even though it takes time for collaboration to yield a research article with particular PACS codes, both collaboration and PACS codes are assessed from publications, and so predicting future collaborations from present PACS codes indicates the likelihood that a similar publishing focus at time $t$ will attract like-minded physicists to work together at time $t$ +1. Conversely, predicting future PACS codes from current collaborations indicates the likelihood that collaborations with a partner institution at time $t$ anticipates a convergence in publishing focus at time $t$ +1. When we perform these regressions between each pair of universities (Figure 4A), we discover the degree to which social connection leads to convergence, and its converse. When we perform them for each university and all other universities (Figure 4B), we unearth the extent to which social connection leads to cultural contraction. Figures 4A and B summarize all of these Granger causal regressions. The number decorating each arrow in the graph represents the percentage of significant Granger causal regression estimates that post a positive influence.

---

Figure 4 about here

---

From these figures, we see that the apparent causality between social and cultural distances is bidirectional, suggesting that both socialization and selection are driving the attention of universities' physicists. Nevertheless, present collaborative distance drives future cultural or topical distance more consistently than its converse. 67% of all regressions predicting next year's topic distance from each pair of universities' current social distance post a positive effect, compared with only 57% that do so in the opposite direction. By contrast, 89% of all regressions predicting next year's topic distance from current global collaborations between a university and all others post a positive effect, compared with 79% from collaboration to shared focus. These patterns provide moderate evidence that social connection leads to cultural convergence between pairs of universities, but strong evidence that more global social connections lead to cultural contraction.

When we decompose the hyperbolic distance between pairs of universities, and between each university and all others, into the difference between their angle ("divergence") and radius ("hierarchy"), we find that hierarchical distance in the social or cultural domain is always a more consistent predictor of hyperbolic distance in the other domain than angular distance. This suggests a more consistent sensitivity of social and cultural distances to differences in the centrality of a university's physicists in the alternative space. Greater centrality provides universities and their physicists with a wider menu of options, and less centrality gives them far fewer choices regarding what they can do or where they can go next. As a result, differences in social centrality more consistently magnify differences in cultural distance, and the converse, than differences in angle, which represents a university's specific topical focus or collaboration cluster. This provides substantial support for the importance of using of a hyperbolic representation that highlights hierarchy for the social and cultural systems characterized by it.

In appendix Figure A5, we validate the relationship between social connection and cultural contraction with standard network and information theory measures. In the inset of the right panel, we see that when an institution's clustering coefficient in the network goes up, their entropy over PACS codes goes down. This suggests with greater density comes less (individual) diversity. We note that the hyperbolic measures between pairs of universities and between each university and all others provides more dispositive and descriptive information regarding social connection and contraction.

**Discussion**

We live in an age of social hyper-connectivity. From Facebook and the rise of social media, to the drop in transportation costs that facilitate social gathering, meeting, and mixing, there are more opportunities to connect and remain weakly connected with a wider range of persons than ever before. In the sociological and organizational literatures, the process of bridging social (Burt 2009) and cultural holes (Pachucki and Breiger 2010) has been linked with a wide range of individual, organizational and city-level benefits through increased access to cultural information, and opportunities for strategic and creative recombination. Individual benefits incentivize strategic networking, enabling entrepreneurs to engage in arbitrage across communities and markets (Burt 2009). Scientists and engineers similarly strive to place their ideas and inventions at the center of complex networks in order to cement their continuing influence and importance (Latour 1987). All of these forces result in network extension and expansion, which shortens the social distance between ideas, symbols and other cultural forms, ultimately placing them in competition and resulting in rising rates of cultural extinction. This cycle of social exploration and cultural exploitation characterizes processes of colonialism, globalization, and capitalism, resulting in cultural creativity, convergence, and sometimes collapse. The process is manifest in the extinction of indigenous languages and cultural forms.

In this paper, we built an information theoretic foundation for understanding the relationship between social connection and cultural contraction. Greater social connection unleashes multi-party communication and cultural exchange, which correlates cultural expressions in future communication. These correlations can be positive, when parties agree, or negative, when parties disagree but nevertheless entrain one another into polarizing shared focus (e.g., pro-life versus pro-choice).

We argued that social and cultural worlds could be powerfully represented with hyperbolic geometry to simultaneously reveal the hierarchy and dispersion of social or cultural forms within them. We discussed recent advances in neural networks that have made these hyperbolic embeddings practical, and we rendered the university collaboration network of 21$^{st}$ Century physics traced by the corpus of the American Physical Society within a two dimensional Poincaré disk. We comparably represented the epistemic culture of 21$^{st}$ Century physics. We found that the social connection of academic physics was becoming more dense over time, and that universities tended to sponsor research that increasingly traded off between particle and nuclear physics, on the one hand, and condensed matter, on the other. We ran Granger causal analyses for pairs of universities and found that social collaborations more consistently predict future cultural shifts than the converse. We ran the same analyses for each university in their collaborations with all other universities, finding that greater external collaboration consistently leads to future convergences with globally popular topics. This was supported by a standard analysis of network density and information theoretic diversity.

There are several limitations of our analysis. First, our case of modern physics may not be perceived as representing the fluidity and spontaneity of other "cultures". Nevertheless, we argue that it represents a conservative case, given the strong rewards for topical consistency in academic promotion. Second, our short time frame, 2002-2011, does not allow us to observe long-term change and influence between social and cultural spheres. Third, by calculating statistics like percentages over many estimated models (see Figure 4), we implicitly avoid taking into account the social and cultural dependencies between cases. Hyperbolic distances between university pairs in the collaboration and PACS code networks were constructed in context of all other collaborations and published PACS codes and so their measurements and associated cases are not truly independent from one another. Despite these limitations, we believe that our findings shed substantial light on the cultural costs of connection.

The relationship between social connection and cultural contraction poses a deep paradox for modern, multicultural societies that value diversity. Extensive research in social, behavioral and organizational science documents the largely positive effect social and cultural diversity exerts on the collective production of information, goods and services (Joshi and Roh 2009; Page 2008). Individuals from culturally distinct groups embody diverse cultural perspectives and cognitive resources that combine to produce solutions and designs that outperform those from homogeneous groups (Mannix and Neale 2005; Woolley et al. 2010; Hong and Page 2004; Nielsen 2012). Collaborations between inventors from distinct social groups result in more creative patents (Fleming, Mingo, and Chen 2007), scientific teams representing distinct disciplines produce more highly cited papers (Wuchty, Jones, and Uzzi 2007), gender diversity broadens the questions scientists ask (Nielsen 2012), and political diversity leads to improved political pages on Wikipedia (Shi et al. 2017). Beyond performance, some enjoy a taste for difference or heterophily (Lazarsfeld, Merton, and Others 1954). But if and when cultural differences fuse through social integration, they become less different. And when differences erode, so do the performance benefits that come from them. Social connection consumes cultural diversity.

Only by considering the relationship between society and culture on both short and long time scales can we clearly see the essential tension between the enjoyment and protection of cultural difference. Cultural differences and cultural forms can be generated through isolation and specialization, but as with indigenous languages like Wichita, Mandan and Yurok, or megafauna like the Tasmanian tiger, Chinese Baiji dolphin, or North American heath hen, all of which irreversibly disappeared within the last hundred years, it is both faster and easier to extinguish cultural and biological forms than to evolve new ones. We argue for greater incorporation of long-run cultural dynamics into theories of social connection, integration and change to insure the long-run benefits of diversity.

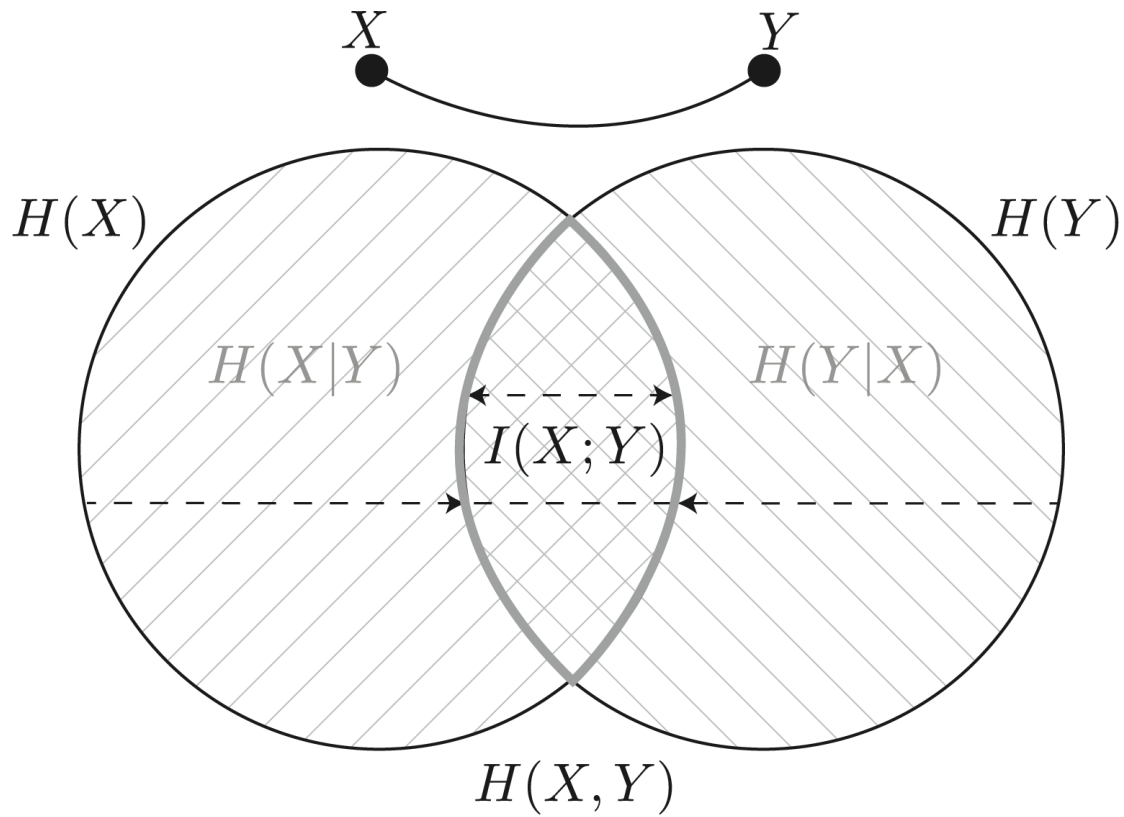
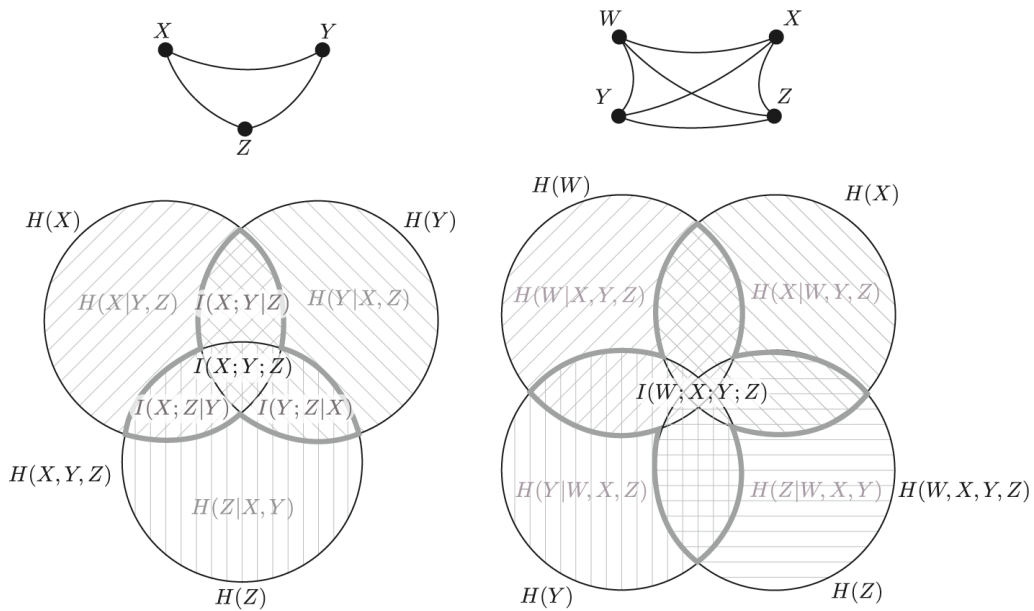

**Figure 1. Venn diagrams for various information measures calculated for culturally correlated actors W, X, Y and Z.** For the simple {X, Y} network, H(X) constitutes the entropy or average information associated with X's cultural forms and H(Y) the entropy of Y's cultural forms. As their relationship and similarity increases, their mutual information I(X;Y) rises and their joint entropy H(X,Y)—the total information in the system—shrinks. H(X|Y) is the conditional entropy of X given Y, or the average cultural information or surprise another actor A would experience when communicating with X after having communicated with Y. The graphs and Venn diagrams at the bottom of the figure illustrate the information scenario for clustered networks {X, Y, Z} and {W, X, Y, Z} (with a missing W, Z intersection for the {W,X,Y,Z} network), where the joint entropy shrinks linearly with the union of all multivariate mutual informations (see footnote 3).

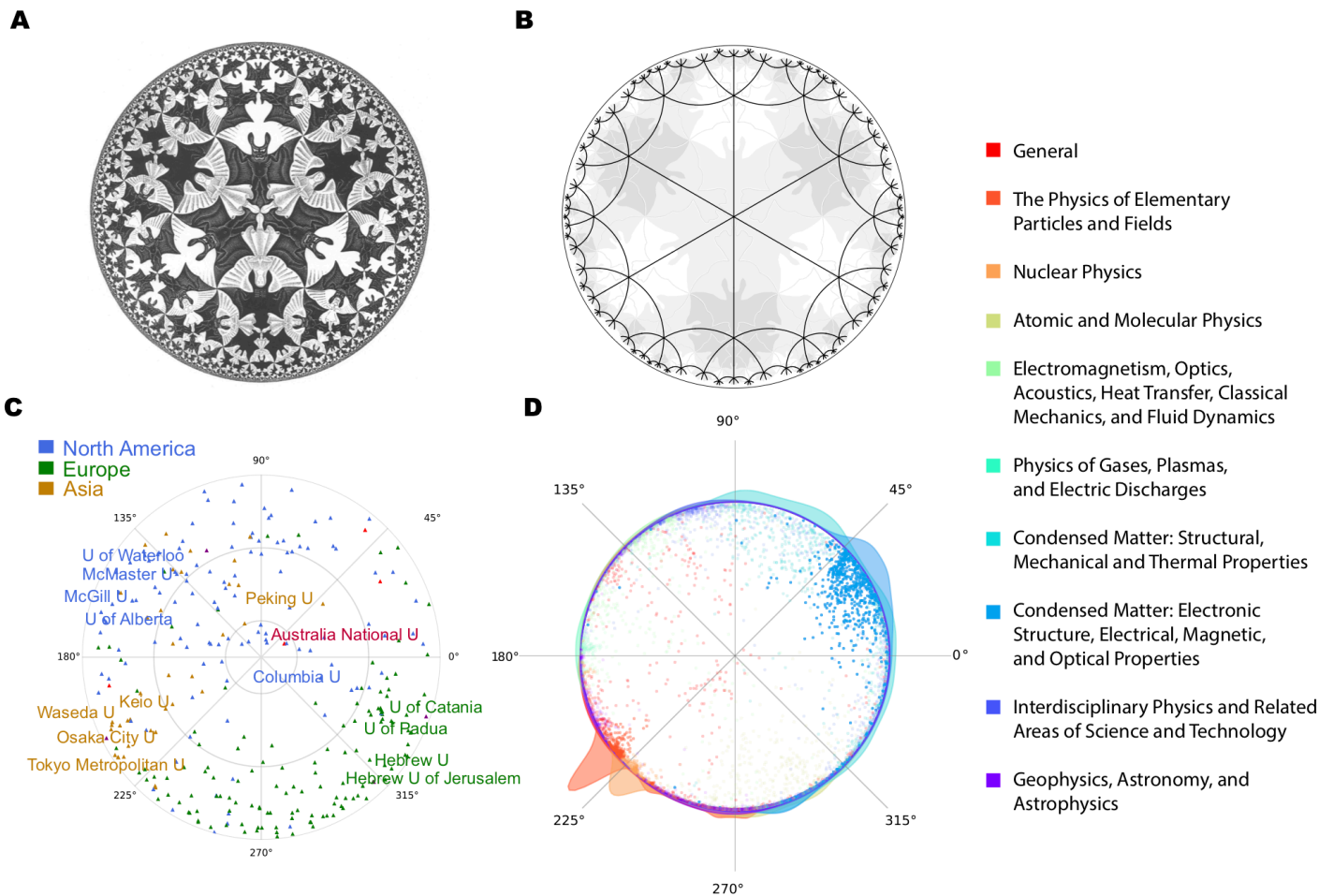

**Figure 2.** Panel A and B present M.C. Escher's print *Circle Limit IV* and the underlying the Poincaré disk model, recited and recreated from (Dunham and Others 2009). The space is getting exponentially denser as the radius increases. Panel C shows the hyperbolic embedding of the institution collaboration network in 2011 using the the Poincaré disk model. The displayed 290 institutions (triangles) are colored by regions: blue for North America, green for Europe, orange for Asia, red for Australia, and purple for South America. Panel D shows the hyperbolic embedding of the aggregated PACS code co-occurrence network using the the Poincaré disk model. The analyzed 5,819 PACS codes (dots) are colored by the 10 subfields they belong to. The Gaussian kernel density estimation is used to visualize the concentration of the angles of PACS codes in each subfield (see Figure A4 for details). To compare across subfields, we also rescale the estimated distributions such that the area covered by the distribution curve is not unity, but proportional to the total number of papers published by 290 institutions in this subfield from 2002-2011.

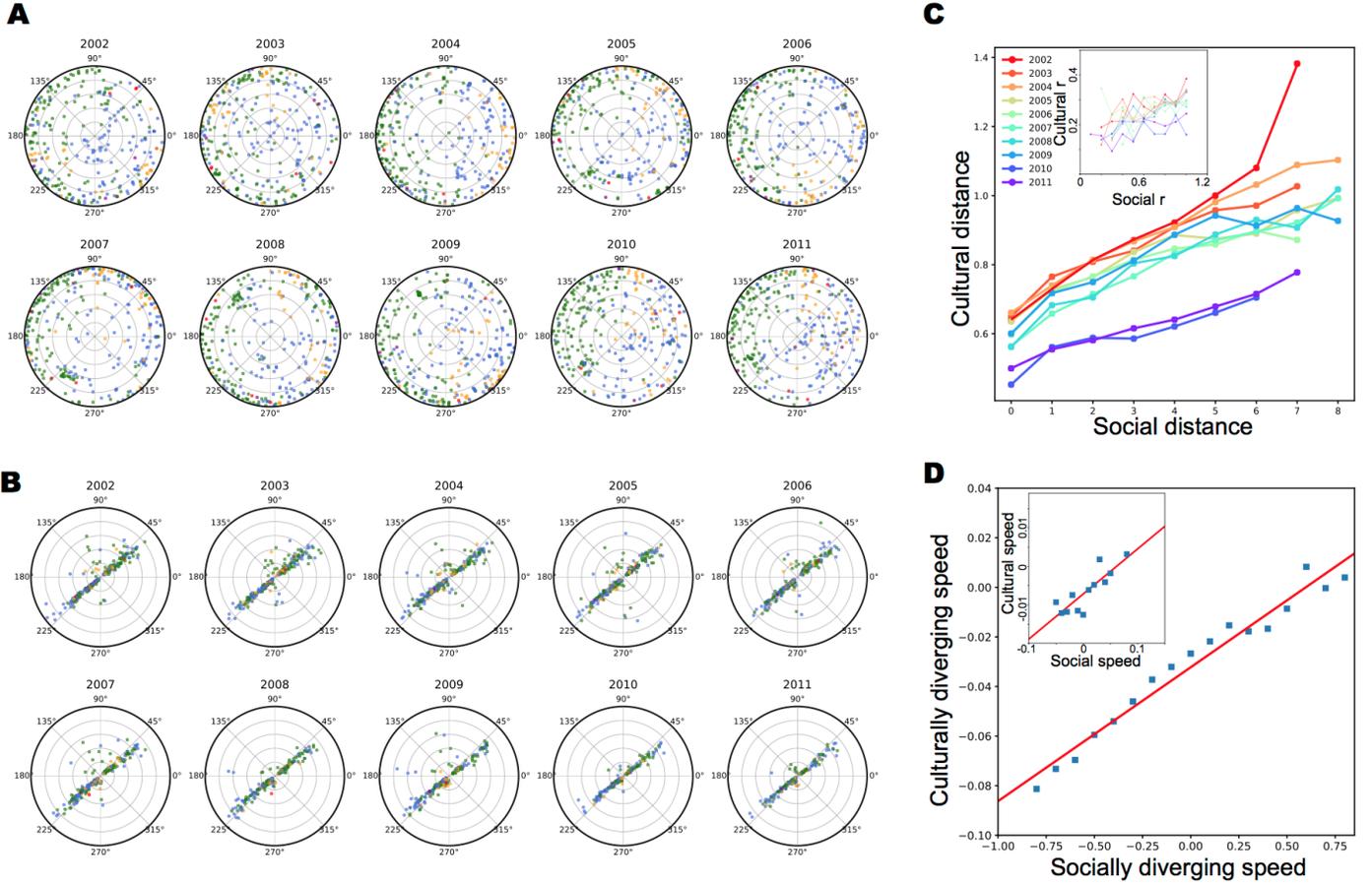

**Figure 3.** (A) The hyperbolic embedding of annual university physics collaboration networks (2002-2011). We embed each annual collaboration network into a hyperbolic space and project them on Poincaré disks, such that each institution $i$ has two parameters in the hyperbolic social space, angles and radius . The universities (dots) are colored by region, blue for North America, green for Europe, orange for Asia, red for Australia, and purple for South America. (B) Annual projection of institutions in the hyperbolic space of physics topics. We embed the PACS code co-occurrence network into Poincaré disks as shown in Figure 2D, and then calculate the position of each university in a given year from the position of their PACS codes, such that each institution $i$ has two parameters in the culture space, angles and radius . The manner by which we we aggregate the PACS code parameters is illustrated in Figure A4. (C) The correlation between social and cultural distances. We calculate the hyperbolic distance between every pair of universities in both Social and Cultural space as and for all pairs of institutions in each year, then plot against across years (data is binned). The inset shows the correlation between social and culture hyperbolic distances at the university level, i.e., we calculate the average hyperbolic social and culture distances from one institution to all the other institutions $\frac{1}{ }\sum_{=1}$ and $\frac{1}{ }\sum_{=1}$ , and plot them against on another. (D) The correlation between the social and cultural time series, social distance ( ) and cultural distance ( ). For each pair of institutions, we use an OLS regression to fit the slopes (social diverging speed) of ( ) against year $t$ and $\beta$ (culturally diverging speed) of ( ) against years $t$. The large, positive slopes imply that two institutions are moving toward and away from each other in the corresponding spaces. We find that and $\beta$ are positively correlated, with a Pearson correlation coefficient equal to 0.98 (*P-value* < 0.001). Similarly, in the inset we show that the slopes of the time series ( ) and ( ) are also correlated, with a Pearson correlation coefficient equals 0.85 (*P-value* < 0.001).

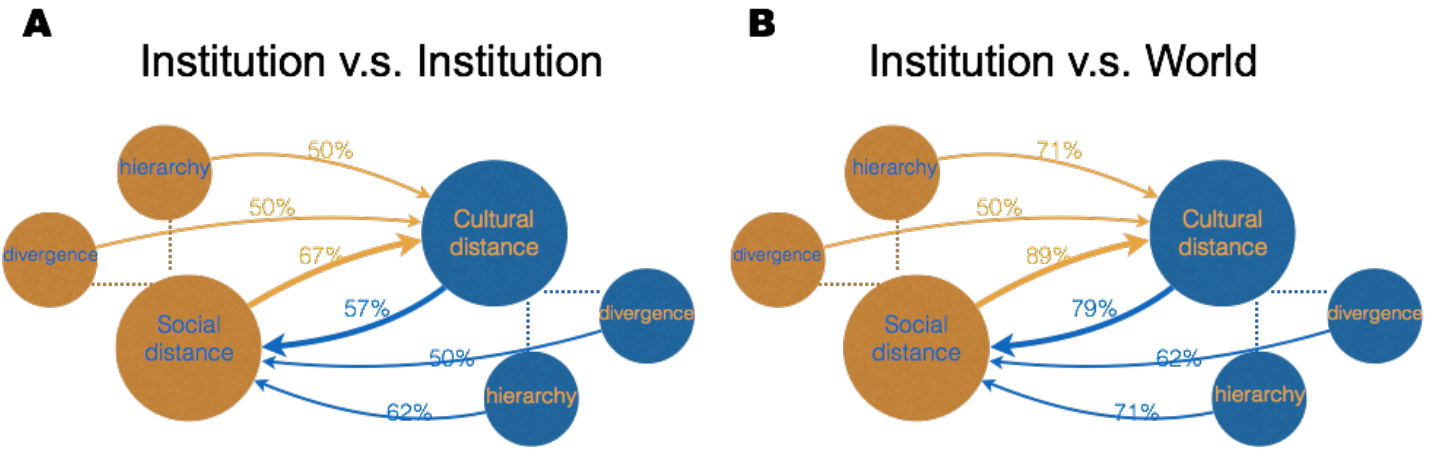

**Figure 4.** We use Granger causality regression to explore whether the social distance of institutions predicts their cultural distance in future and vice versa. In addition, as the positions of institutions and the distance between them always involves two factors, angle ("divergence") and radius ("hierarchy"), we can further investigate which factor contribute more in the prediction. We investigate the dynamics between pairwise institutions ("institution vs institution", panel A), and also aggregate the results to explore how single institutions interact with all other institutions ("institution vs world", panel B). In both panels, arrows point from the predictors to the predicted variables. The values associated with arrows were the percentages of positive coefficients in the Granger causality regression, which quantifies the predicting power in the positive direction.

# Appendix

In Figure A1, we graph the institution collaboration network over time, with all 290 institutions (dots) colored by region: blue for North America, green for Europe, orange for Asia, red for Australia, and purple for South America. Edges represent the co-authoring relationship between scholars from each institution. Between 2002 and 2011, the number of edges increased from 3,331 to 7,647, and the total weight increased from 29,117 to 94,336 over the same period.

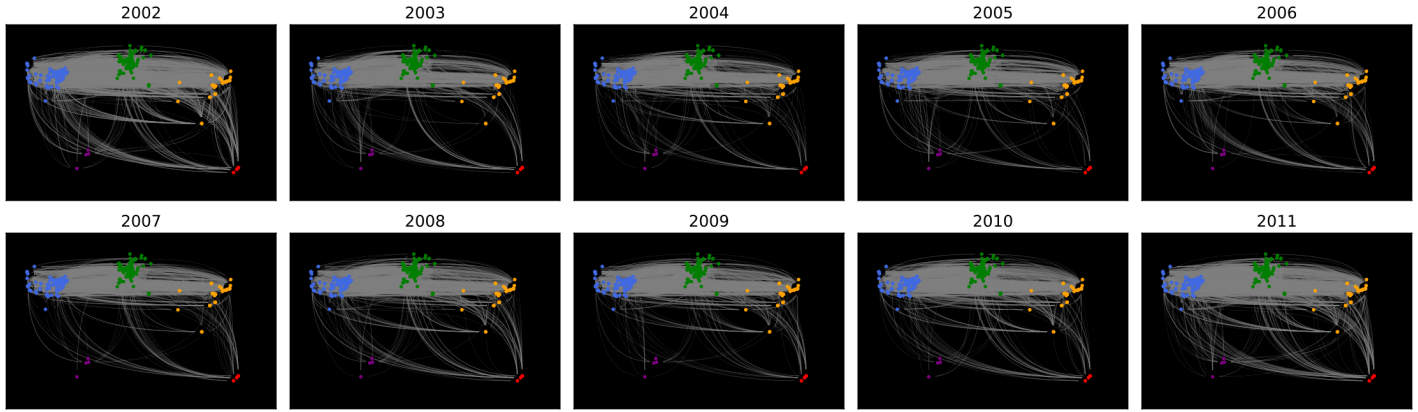

**Figure A1. Institution collaboration network, 2002-2011.**

In Figure A2 we illustrate the movement of institutions in collaboration space over time. We pick Columbia University as the focal institution (the red triangle) and investigate the temporal evolution of its distance to two collaborating institutions, the University of Notre Dame (the orange rectangle) and University of Southern California (the blue rectangle). Over time, the University of Notre Dame moves towards Columbia University, whereas the University of Southern California moves away from it.

We also highlight the top ten institutions closest to in Columbia University in 2002 (purple dots) and their position in the following years to provide visual guidance for distance in the hyperbolic space, which differ substantially from 2D Euclidean distance.

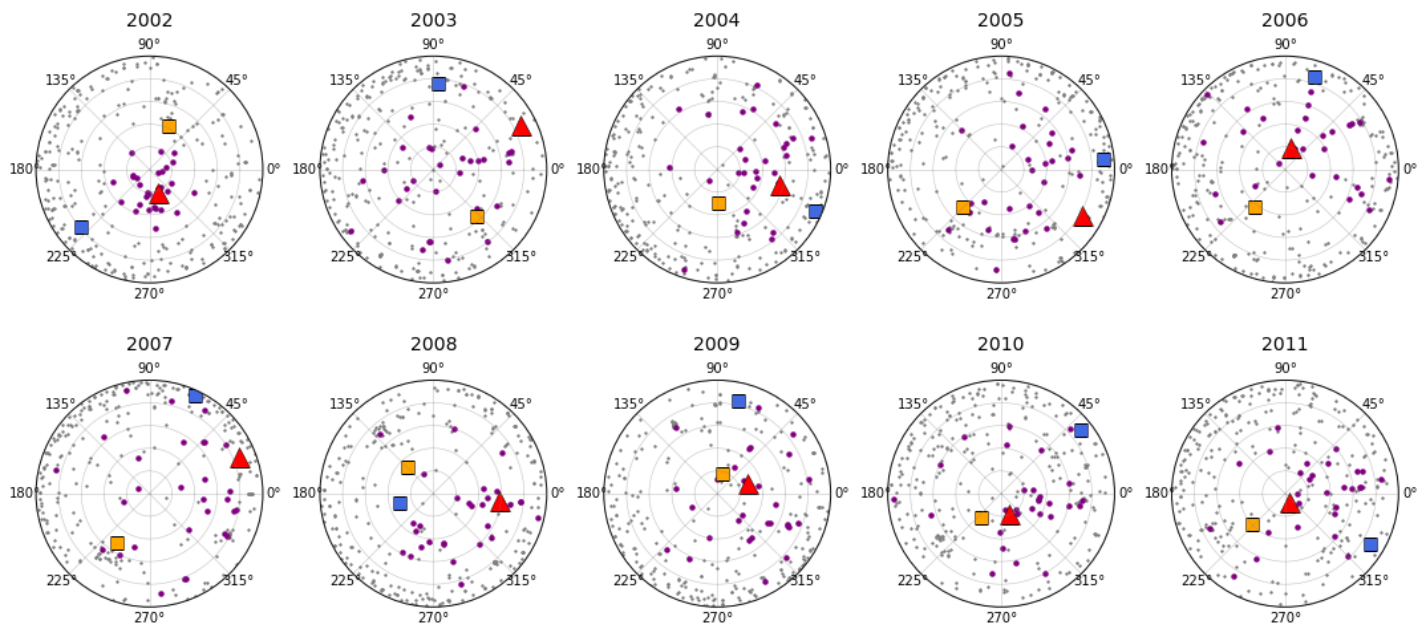

**Figure A2. Movement of institutions in collaboration space, 2002-2011.**

In Figure A3, we illustrate the movement of institutions in the cultural space of physics over time. We constructed the culture space as shown in Figure 2B, and display all 5,819 PACS codes in the panels above as gray dots. Next, we calculated the representative locations of Columbia (red triangle), Notre Dame (orange rectangle) and University of South California (blue rectangle), as shown in Figure 2B, but within the culture space based on the positions of their PACS codes using the method introduced in Figure A4. We also fitted Gaussian kernel density curves to show the concentrate and spread of PACS codes for each of these three institutions.

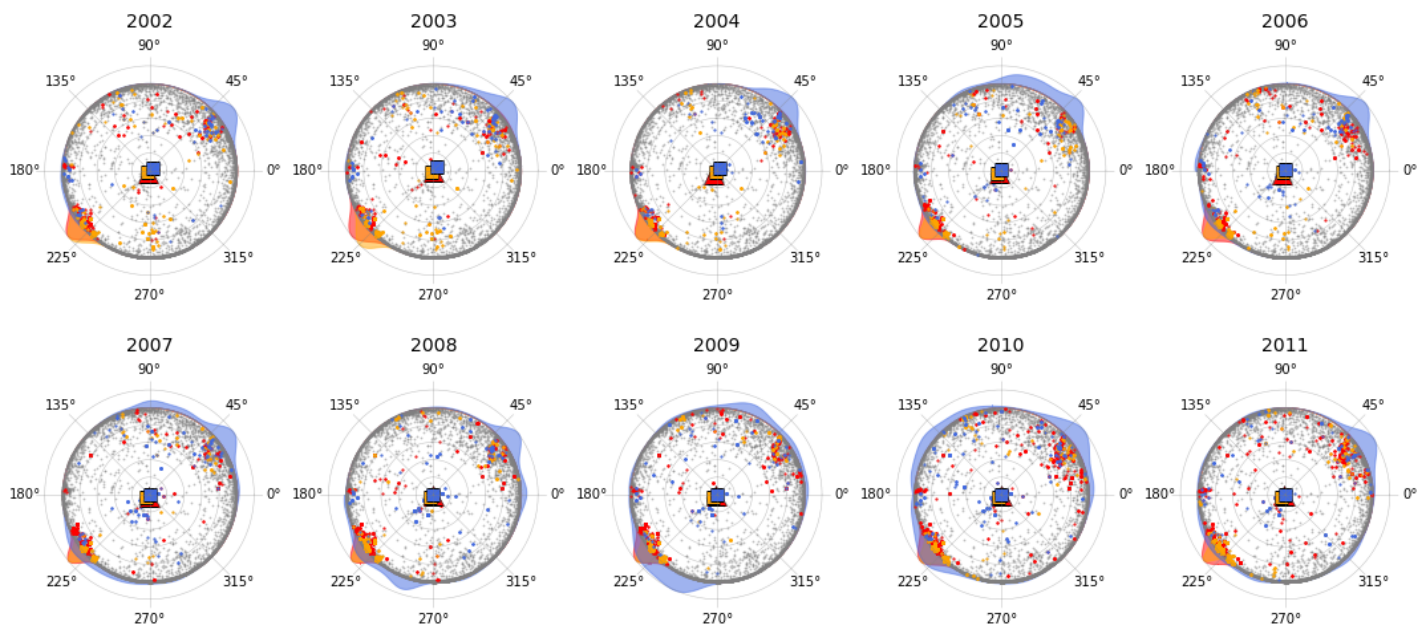

**Figure A3. Movement of institutions in the "cultural space" of physics topics, 2002-2011.**

In Figure A4, we illustrate the empirical distribution of angles and radius for PACS codes from Figure 1D. Each row shows the 5,819 PACS codes from one of ten subfields, which are colored following the same scheme as Figure 1D. Due to the size limitations of the figure, abbreviated names of the subfields are used. The first column shows the two estimated Poincaré disk parameters, angles $\theta_i$ (the x-axis) and radius $1 - r_i$ (the y-axis; for figure readability rather than $r_i$) of the *i*th PACS code. The second column shows the empirical distribution of angles $\theta_i$ and corresponding Gaussian kernel density estimation. The red lines show the peak of the angle distribution we select to represent the summary research direction of subfields or institutions based on their PACS codes. The third column shows the empirical distributions of radius $1 - r_i$. We found that due to the asymmetric property of the hyperbolic space, the values of most $r_i$ approach 1. As such, neither mean nor median can characterize the difference in hierarchy between two universities or subfields. Therefore we take the extreme values by calculating the average of the ten smallest $r_i$ to represent the university or subfield hierarchy.

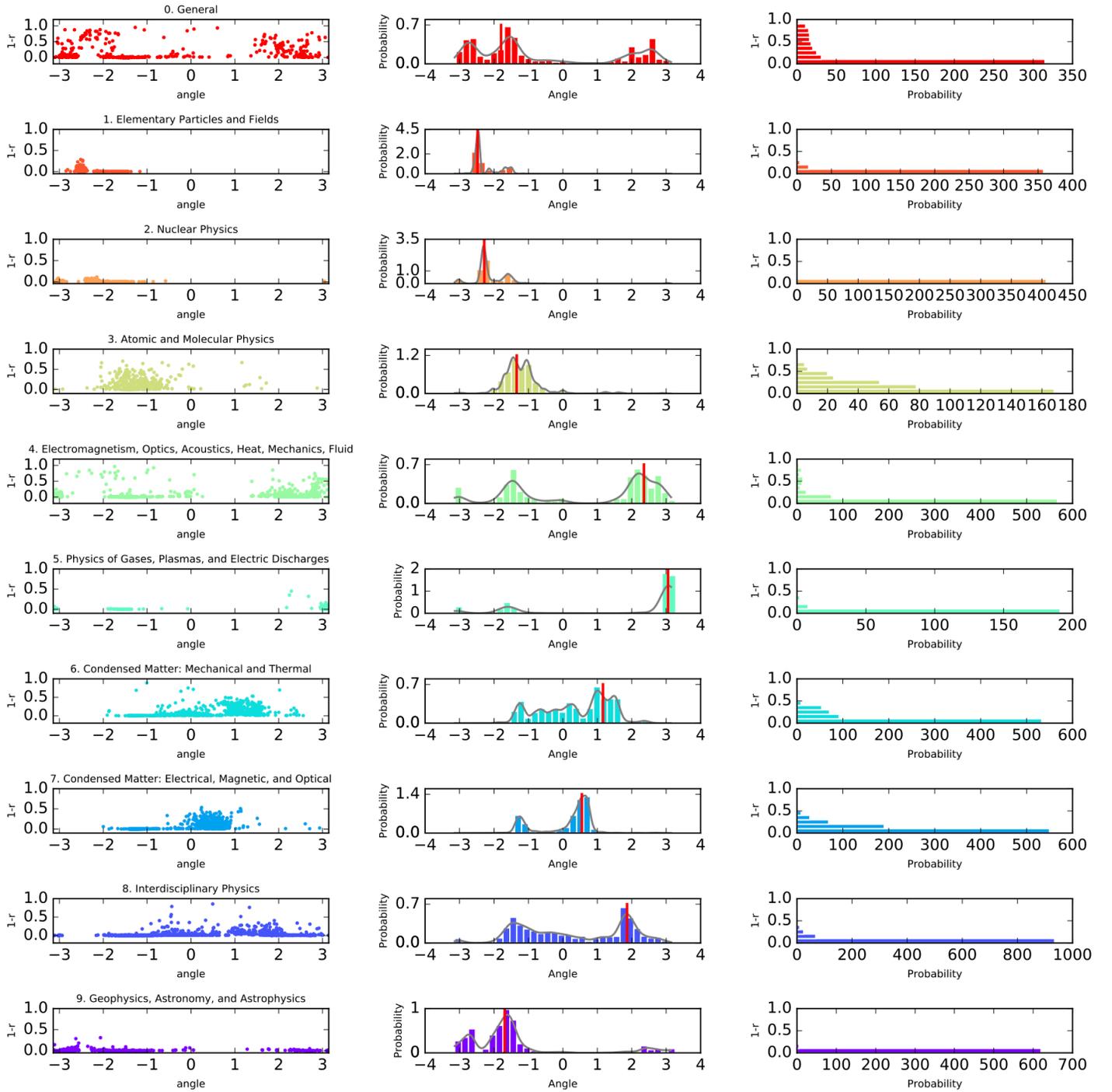

**Figure A4. Illustration of the empirical distribution of angles and radius for all PACS codes from Figure 1D.**

After we constructed a university collaboration network, we also measured the density of social collaborative connections. Following convention, we define social density as the ratio of observed network edges to the maximum number possible (Wasserman and Faust 1994). We denote the set of weighted edges connecting the *i*th node into the global institutional collaboration network as     , which can be further separated into      =      +      , where        denotes self loops and          edges connecting node *i* with other nodes (Palla, Barabási, and Vicsek 2007). For institution *i*, we defined a weighted clustering coefficient:

$$SDC_i = \sum (|N_i \cap N_k| \cdot |N_i \cap N_h| \cdot |N_k \cap N_h|)^{1/3} / (max(w) * |N_i| * (|N_i| - 1))$$

where *k* and *h* include all unique pairs of institutions that constitute closed triplets with *i* and *max(w)* represents the maximum weight of *i*'s ego network. We use $SDC$ to measure social density as an alternative to measures defined in the hyperbolic social space (Figure 3). We also calculated the entropy of PACS codes as a measure of cultural diversity. In Figure A5 we find that social density and cultural hierarchy are positively correlated and that social density and cultural diversity are negatively correlated. Therefore, our expectation that the broad association between social interaction and cultural contraction appears to hold for 21st Century Physics.

    If position in cultural hierarchy is defined as radius $r$ and cultural diversity $H$ as the entropy over their PACS code angles on the Poincaré disk, then the two main figures show the temporal evolution of $SDC$, $r$, and $H$ for two extreme cases. The left panel shows how social density ($SDC$, the left y-axis) and cultural hierarchy ($r$, the right y-axis) move in the same direction. More specifically, both $SDC$ (dotted red line) and $r$ (solid red line) for the University of Innsbruck decreases over time. We plot these as ranks and not original values and also adjust the scale of ranks to provide visual guidance for comparing trends. By contrast, both $SDC$ (dotted blue line) and $r$ (solid blue line) for the Naval Research Lab increase over time. The right panel shows how social density ($SDC$, the left y-axis) and cultural diversity ($H$, the right y-axis) go in the opposite direction. The $SDC$ for the University of Innsbruck decreases over time (dotted red line), whereas its PACS code entropy increases (solid red line). By contrast, the $SDC$ of Johns Hopkins University increases over time (dotted blue line), but its PACS code entropy decreases (solid blue line). The two insets confirm that the full sample social density and cultural hierarchy are positively correlated and the full sample social density and cultural diversity are negatively correlated.

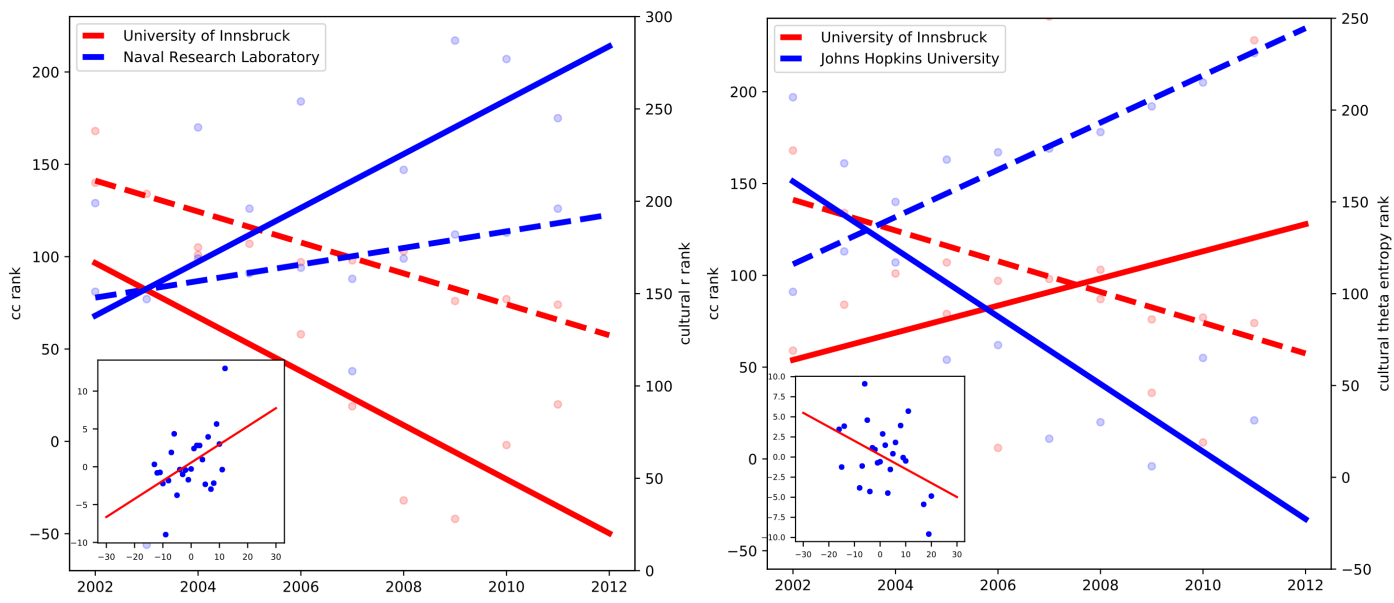

**Figure A5. Correlation between social density, cultural hierarchy, and cultural diversity.**

Figure A6. Hyperbolic embedding tags from 47,282 questions in Physics Stack Exchange (https://physics.stackexchange.com/).